\documentclass[sigconf,final]{acmart}

\AtBeginDocument{%
	}

\copyrightyear{2022}
\acmYear{2022}
\setcopyright{acmcopyright}\acmConference[ICAIF '22]{3rd ACM International Conference on AI in Finance}{November 2--4, 2022}{New York, NY, USA}
\acmBooktitle{3rd ACM International Conference on AI in Finance (ICAIF '22), November 2--4, 2022, New York, NY, USA}
\acmPrice{15.00}
\acmDOI{10.1145/3533271.3561685}
\acmISBN{978-1-4503-9376-8/22/10}

\usepackage{caption}
\usepackage{subcaption}
\usepackage{dsfont}
\usepackage{csquotes}
\usepackage{array}
\usepackage{soul}

\newcommand{\E}{\mathbb{E}}

\acmSubmissionID{XX}

\begin{document} 
\title{Can maker-taker fees prevent algorithmic cooperation in market making?}

\author{Bingyan Han}
\affiliation{%
  \institution{Univeristy of Michigan, Ann Arbor}
  \country{USA}
}
\email{byhan@umich.edu}

\renewcommand{\shortauthors}{Han}

\begin{abstract}
 In a semi-realistic market simulator, independent reinforcement learning algorithms may facilitate market makers to maintain wide spreads even without communication. This unexpected outcome challenges the current antitrust law framework. We study the effectiveness of maker-taker fee models in preventing cooperation via algorithms. After modeling market making as a repeated general-sum game, we experimentally show that the relation between net transaction costs and maker rebates is not necessarily monotone. Besides an upper bound on taker fees, we may also need a lower bound on maker rebates to destabilize the cooperation. We also consider the taker-maker model and the effects of mid-price volatility, inventory risk, and the number of agents.  
\end{abstract}



\maketitle

\section{Introduction}
Modern security markets have complicated structures and are becoming more fragmented. There are registered national securities exchanges and alternative trading systems as trading platforms. Exchanges face competition to attract broker-dealers for order routing. Broker-dealers usually consider price quality, transaction costs, potential incentives, price impact, and execution speed. In particular, the transaction fee structures play a pivotal role in their decisions.

Previously, exchanges imposed transaction fees on all parties involved in a trade, including liquidity suppliers known as market makers and investors as liquidity takers. Later, some venues started to offer rebates to the makers and charge access fees to the takers. More than half of the national exchanges in the United States currently adopt this ``maker-taker'' fee model. To avoid venues increasing their liquidity rebates by turns, in 2005, the SEC adopted a rule prohibiting exchanges from imposing transaction fees over \$0.0030 per share, which is also an implicit cap on rebates. They chose the \$0.0030 level because it was consistent with business practices at that time.

In academia and industry, there is a long debate over whether this fee cap is reasonable and whether the maker-taker model harms the public interest. To collect relevant data and study the effects of transaction fees, the SEC proposed a transaction fee pilot program in 2018, with test groups and a control group on selected stocks. Some market participants were in favor of the pilot, while major exchanges believed that the pilot would ultimately harm investors. After timely appeals from exchanges, the U.S. Court of Appeals for the D.C. Circuit halted the program and ruled that the SEC overstepped its authority, see the opinions released on June 16, 2020. \footnote{Available at {\scriptsize\url{https://www.cadc.uscourts.gov/internet/opinions.nsf/BE5AD5AD3C0064408525858900537163/$file/19-1042-1847356.pdf}}} 

In the mathematical finance literature, market making has been formulated as a stochastic control problem \citep{avellaneda2008high,gueant2013dealing}. The maker-taker fee model is investigated as a principal-agent problem for one market maker \citep{euch2021optimal} and multiple makers \citep{baldacci2021optimal}. Recent advances in artificial intelligence stimulate the interest in tackling market making with reinforcement learning. Examples include but are not limited to \cite{gordon2017,spooner2018market,nips19MM,ardon2021towards}. As the real market data are expensive and impossible to collect after the appeal on the pilot, simulation becomes an attractive way to investigate the effect of maker-taker fees. Another motivation, however, comes from preventing unexpected consequences by algorithms.  

Presently, market participants are increasingly adopting algorithms in security trading. However, the sophistication and powerfulness of algorithms have also led to another prominent concern on collusion. Algorithms may be sufficiently advanced to learn that it is optimal to collude \citep{ezrachi2016virtual}. Several recent experimental studies \citep{waltman2008q,klein2021autonomous,calvano2020artificial,hansen2021frontiers} suggest that algorithms can learn collusive strategies from scratch, even without human guidance or communication with each other. \cite{calvano2020protecting} further points out the difficulty of preventing algorithmic collusion with the current antitrust laws. For the market-making problem, \cite{xiong2021interactions} finds collusive prices are possible under decentralized multi-agent reinforcement learning without price information sharing. \cite{han2021cooperation} considers Q-learning with Boltzmann selection and proves the convergence to supra-competitive spread levels when agents have no memory.

Current antitrust laws usually use communication between agents to identify collusion, due to the concealment of collusive strategies and the difficulty in evaluating supra-competitive prices relative to marginal costs \citep{calvano2020protecting}. For example, in the odd-eighth scandal \citep{christie1994nasdaq}, \citet[footnote 10]{christie1995policy} mentioned that there was overt harassment of dealers who broke the spreads, detailed in the {\it Los Angeles Times} on October 20, 1994 (pg. 1). In contrast, if algorithms can collude without communication, the regulators lack the tools to stop them. An OECD Secretariat background paper\footnote{Available at {\scriptsize \url{https://www.oecd.org/competition/big-data-bringing-competition-policy-to-the-digital-era.htm}}} mentioned that ``{\it there is no legal basis to attribute liability to a computer engineer for having programmed a machine that eventually `self-learned' to coordinate prices with other machines}''.

In economics and antitrust laws, tacit collusion means supra-competitive prices and punishment schemes for deviation. However, in market making, any punishment is obscure when competitors' actions are unobservable, especially for multiple agents. In this work, we only validate supra-competitive prices and term it algorithmic cooperation. Security markets have unique characteristics due to high-frequency trading. Price adjustments are fast since millions of interactions can happen very quickly. Algorithms may converge and maintain supra-competitive outcomes after a short time. Another difference between e-commerce and market making is that brokers commonly quote on both ask and bid sides as both sellers and buyers. 

What is a suitable maker-taker fee level if we use them to prevent algorithmic cooperation? In this paper, we design a semi-realistic security market and adopt a multi-agent reinforcement learning framework. Simulation is much cheaper and safer than experiments in the real market. We consider multiple market makers quoting on a stock in discrete time steps. Wider spreads increase profits per order but attract fewer trading volumes. There usually exists a prisoner's dilemma. If all makers maintain wide spreads, investors have no choice but to accept the transaction costs. If one of the market makers deviates and improves quotes, he earns more by attracting more orders. However, if all of them do so, the profit drops. The odd-eighth scandal \cite{christie1994nasdaq} shows the possibility of this reward mechanism.

We utilize the classic Q-learning framework since it is easier to interpret the parameters and learning progression. The payoff matrix highly depends on the arrival probability of market orders. Inspired by \cite{baldacci2021optimal,nips19MM,avellaneda2008high}, we incorporate the dependence on limit order book quality and mid-price volatility. With the terminology of Q-learning, bid-ask quotes are encoded as discrete actions and the current market condition is summarized in the state variable. We obtain the following main observations experimentally:
\begin{enumerate}
	\item The maker-taker model can reduce net transaction fees only when investors are sensitive to the net spreads in a trade. Indeed, if investors treat all price levels equally, market makers will quote the highest since they will not receive fewer orders. Moreover, the relation between net spreads and rebates is not necessarily monotone. High maker rebates may increase transaction costs, for example, when makers have already quoted the lowest. Interestingly, besides an upper bound on taker fees, a lower bound on maker rebates is probably also needed. Since tick sizes are discrete, a low rebate may not be attractive enough to narrow the spread. Instead, the investors have to pay the same spread plus an extra fee.
	\item Another method, the taker-maker fee model, compensates the liquidity taker but charges the maker. It has opposite effects on price levels since the makers tend to raise the spreads in this model. However, makers are reluctant to do so if investors are sensitive to transaction costs. Then the compensations reduce the net fees for takers. Besides, this model can improve trading volumes significantly.
	\item We consider the impact of mid-price volatility, inventory risk, and the number of agents. High mid-price volatility facilitates the cooperation between makers to raise spreads. Then the rebates should be large enough to reduce transaction costs. However, we may face a dilemma that the required taker fees are too high, such that the net transaction costs are even close to the no-rebate case. For inventory risk, rebates compensate makers for bearing the risk and providing liquidity. If makers are more inventory-risk averse, the required rebates are higher. For the number of agents, if a trading venue has enough makers, maker-taker fees will increase transaction costs for investors instead of promoting competition between makers. Regulators have to include the number of agents as a factor when evaluating the effectiveness of maker-taker fees.    
\end{enumerate}

The rest of the paper is organized as follows. Section \ref{sec:frame} describes the market-making problem, the maker-taker fee model, and the reinforcement learning implementation. Section \ref{sec:exp} presents the experiments, focusing on the effects of maker-taker fees, taker-maker fees, inventory risk, mid-price volatility, and the number of agents. Section \ref{sec:con} concludes the paper with future questions. The code is publicly available at \url{https://github.com/hanbingyan/MRTR}.      

\section{Framework}\label{sec:frame}
\subsection{Market making}
Suppose there is a dealer market with $N$ market makers (agents) and a single traded security. Transactions happen in discrete time steps, as in \cite{nips19MM}. There are mainly two types of orders in the market. Limit orders set the maximum or minimum price at which a participant is willing to buy or sell. The order is executed only when the price matches. Market orders are executed immediately with the best price available. We assume market makers always use limit orders as liquidity providers and investors use market orders as liquidity takers. 

Market makers provide prices at which they are willing to buy (bid) and sell (ask). Commonly, there is a limit order book (LOB) for the security, which provides reference pricing information. The mid-price $S_t$ at time $t$ is the mean value of the best bid and ask in the LOB. Market makers also quote their bid-ask spreads relative to $S_t$. In this paper, we assume a discrete tick size. Consider $K$ price levels on one side since prices far from the mid-point are usually not executed. At time $t$, the market maker $i$ is willing to sell at most $a_i(t, k)$ shares with the price of $S_t + \delta_k$. Similarly, he is willing to buy at most $b_i(t, k)$ shares with the price of  $S_t - \delta_k$. $\{ \delta_k \}^K_{k=1}$ are multiples of the tick size in ascending order. We refer to $a_i(t) := (a_i(t, 1), ..., a_i(t, K))$ as the ask curve of the maker $i$. Define the counterpart for the bid side in the same way. The maker can choose an action $c_i(t)$ to control the shape of bid-ask curves $(a_i(t), b_i(t))$. We assume the action space is finite and discrete. 

For later use, aggregate ask quotes of all agents in one period as the following vectors
\begin{equation*}
	A(t) = (..., A(t, k), ...) := \big( ..., \sum^N_{i=1} a_i(t, k), ... \big),
\end{equation*}  
which is the volume available on the ask side for every price level. In the following presentation, we mainly focus on notations of the ask side for simplicity. The bid side is defined in a similar way.

We assume the size of market orders per period for each side follows a binomial distribution with size $M$ and probability $p$. $M$ represents the largest number of market orders available per period on one side. $p$ is the arrival probability of a single order. There exist numerous models for market order arrival intensities, such as in \cite{avellaneda2008high,baldacci2021optimal}. Crucially, the arrival probabilities of market orders should depend on the market liquidity provided by makers. In period $t$, define the arrival probability of a market order on the ask side as $p = p(A(t)) := e^{- f(A(t))}$. Similarly, bid side arrival probability is $p(B(t)) := e^{- f(B(t))}$. We use the same deterministic function $f(\cdot)$ here only for simplicity. Our framework allows a general $f$ to capture some stylized facts in the market. In this paper, we consider $p(\cdot)$ as a decreasing function of the ratio between the spread and the mid-price volatility $\sigma$, specified as
\begin{equation}\label{eq:arrprob}
	f(A(t)) = \frac{1}{\sigma} \sum^K_{k=1} w_k \frac{A(t, k)}{|A(t)|},
\end{equation}
where $|A(t)| := \sum^K_{k=1} A(t, k)$ denotes the total volume. $w_k$ is the weighting factor increasing in price level $k$. We use the percentage of orders at each level to unify the magnitude.

When a market order of size $m$ arrives, it will be executed as follows. If $m$ is large enough to match all outstanding limit orders, we assume any extra market orders will be canceled or revised for the next period. If $m$ is small, denote $k^*:= \inf \{ k: \sum^{k}_{j=1} A(t, j) > m \}$ as the lowest spread level that can fill all the market orders. Then limit orders at level $\{1, ..., k^*-1\}$ are matched. For the remaining market order of size $m - \sum^{k^*-1}_{j=1} A(t, j)$, suppose it will be split to makers proportionally to their quotes $\{a_i(t, k^*)\}^{N}_{i=1}$ at level $k^*$. After this period, assume all remaining limit orders are revised with new prices unless filled. This assumption is motivated by the fact that a considerable amount of limit orders are finally revised or canceled in practice.  

Denote $g_{i, a}(t, k)$ as the orders received by the maker $i$ in period $t$ for price level $k$ on the ask side. Then the reward on the ask side is
\begin{equation}
	r_{i, a}(t) = \sum^{K}_{j = 1} g_{i, a}(t, j) (\delta_j + \beta). 
\end{equation}
Besides the quoted spread $\delta$, constant $\beta$ is the maker rebate per share for liquidity providers. For simplicity, we suppose it is a constant independent of time, price levels, and agents. In contrast, the customer quoting market orders has to pay an access fee $\eta$ per share as liquidity takers. Thus, the net ask price at level $k$ is $S_t + \delta_k + \eta$. The access fee $\eta$ is usually higher than the maker rebate $\beta$. The difference $\eta - \beta$ is the exchanges' profit.  

Besides the profits from bid-ask spreads, market makers bear the risk from inventory in the stock, due to mid-price fluctuations. Denote the accumulated inventory of agent $i$ after trades at time $t$ as $y_i(t)$. The calculation of the inventory is straightforward. If the market maker $i$ receives $a$ ask orders and $b$ bid orders in period $t$, then the accumulated inventory is $y_i(t) = y_i(t-1) + b - a$. Let $y_i(-1) = 0$ by convention.

In summary, the total reward $r_i(t)$ in period $t$ has three parts
\begin{equation*}
	r_i(t) : = r_{i, a}(t) + r_{i, b}(t) - \xi (y_i(t) - y_i(t-1))^2.
\end{equation*}

We deduct the third quadratic cost from the profits to model the inventory risk. Constant $\xi$ represents the aversion to inventory risk and is assumed to be the same for all market makers for simplicity.

For later use, denote $C(t) := (c_1(t), ..., c_N(t))$ as all the actions in period $t$, controlling the distribution on the ask and bid side. Based on the actions, the conditional expectation of $r_i(t)$ is
\begin{align*}
	\E [r_i (t)|C(t)] =&  \E[r_{i, a}(t)|A(t)] + \E[r_{i, b}(t)| B(t)] \\
								&- \xi  \E[(y_i(t) - y_i(t-1))^2| C(t)].
\end{align*}

Consider the market-making game for one round first. We need some game theory concepts to describe the properties of policies \citep{Nowe12}. By dropping the time script under the one-period setting, we have static payoff matrices for market makers. The game in one period becomes a matrix game. A random strategy, also called a {\it mixed} strategy, is a probability distribution of actions. Deterministic strategies are special cases that assign probability one to a single action and are also called {\it pured}. See \cite{Nowe12} for a survey on related concepts. In a Nash equilibrium, each agent acts the best response to other agents' choices.

\begin{definition}\label{def:nash}
	$C^* = (c^*_1, ..., c^*_N)$ is a pure Nash equilibrium if
	\begin{equation}
		\E [r_i | C^*] \geq \E [r_i | C^*_{-i}, c_i]
	\end{equation}
	for all agents $i$ and action $c_i$, where $C^*_{-i}$ is $C^*$ except $c^*_i$.
\end{definition}
In other words, if all other agents $j$, with $j \neq i$, choose $c^*_j$, then it is optimal for agent $i$ to choose $c^*_i$. An arbitrary matrix game may not have a pure Nash equilibrium, even for a two-agent case. However, a two-agent general-sum game always has a mixed Nash equilibrium \citep{Nowe12}. Besides, a matrix game may also have multiple pure Nash equilibria. 

Monopoly inhibits the competition and dampens the efficacy of a market system. Almost all countries have antitrust laws to punish monopoly and collusive behaviors. As pointed out by \cite{calvano2020protecting}, the current law uses the communication between agents as pieces of evidence to identify the collusive pricing. However, if agents can establish supra-competitive strategies with algorithms independently, i.e., without any form of communication, then the law cannot stop them. It raises concerns about algorithmic collusion. From the economic and legal perspective, collusion not only means supra-competitive prices but also involves punishments for deviation from supra-competitive outcomes. In this paper, we only focus on the supra-competitive prices and term it algorithmic cooperation. Mathematically,
\begin{definition}\label{def:coop}
	$\hat{C}^* = (\hat{c}^*_1, ..., \hat{c}^*_N)$  is a cooperative strategy if it maximizes $\E [ \sum^N_{i=1} r_i | C]$, the joint profits of all agents.
\end{definition}         
It is optimal for all agents to select a cooperative strategy $\hat{C}^*$. However, if agents have the motivation to cheat and deviate from $\hat{C}^*$, they may finally result in a Nash equilibrium with much lower profits. 

When extending to a multi-period setting, each agent aims to maximize its cumulative profit $\E[ \sum^\infty_{t=0} \gamma^t r_i(t)]$. $\gamma$ is a constant discount factor. When market makers quote new bid-ask spreads every period, the market-making problem becomes a repeated game. 

\subsection{Reinforcement learning}
Since communication between agents is forbidden, agents do not know others' selections and rewards. If all makers train their market-making algorithms independently, do they always converge to a Nash equilibrium? To answer this question, we adopt independent reinforcement learning (InRL) in this work. Although InRL may not be the exact algorithm implemented in the marketplace, it captures the essence of market making and provides an example to explore. There exist many variants of RL algorithms. We intentionally choose a parsimonious design usually known as the tabular Q-learning. It is a popular algorithm with many applications in the repeated game and is also easier to interpret the parameters. We focus on financial interpretation instead of pursuing more advanced algorithms. 

Reinforcement learning incorporates states to reflect the current information known by agents \citep{watkins1992q,gordon2017,calvano2020artificial}. Suppose market makers observe a common discrete state variable $v(t)$. It can contain the description of the LOB and market liquidity. Define the optimal action-value function $q^*_i(v, c_i)$ for agent $i$ as the maximum expected payoff achievable by following any market making policy $\pi_i$, after observing state $v$ and then taking some action $c_i$,
\begin{equation}\label{Eq:def-Q}
	q^*_i(v, c_i) := \max_{\pi_i} \E \Big[ \sum_{t = 0}^\infty \gamma^t r_i (t) \Big| v, c_i, \pi_i \Big]. 
\end{equation}
Intuitively speaking, this function measures the quality of action $c_i$ under state $v$. Denote any policy that achieves the maximum in \eqref{Eq:def-Q} as $\pi^*_i$. We highlight that the state $v$ only contains the information of LOB, but not the profits, actions, and policies of the competitors. Crucially, $q^*_i(v, c_i)$ satisfies the Bellman equation
\begin{equation}\label{Eq:Bellman}
	q^*_i(v, c_i) = \E \Big[ r_i + \gamma \max_{c'_i} q^*_i(v', c'_i) \Big| v, c_i\Big],
\end{equation}
where $r_i$ is the reward for agent $i$ in one period and $v'$ is the next state observed after taking action $c_i$ under state $v$.

Independent learners treat other agents as part of the environment and use matrices \citep{watkins1992q} or neutral networks \citep{mnih2015human} to learn $q^*_i$. The algorithm faces a trade-off between experimenting with actions that are currently suboptimal (exploration) and continuing to learn the information already obtained (exploitation). Therefore, $\varepsilon$-greedy policy is introduced to follow the current greedy policy with probability $1 - \varepsilon_t$ and a purely random action with probability $\varepsilon_t$. We consider a  time-declining exploration rate, exogenously set as
\begin{equation}
	\varepsilon_t = e^{- \mu t},
\end{equation}
with parameter $\mu > 0$ controlling the time-declining speed of random exploration.

To approximate the unknown $Q$-value function, each agent maintains a separate function $q_i(v, c_i)$. Starting from an arbitrary initial $q_{i, 0}$, Q-learning updates $Q$-values by
\begin{align}
	q_{i, t+1} (v(t), c_i(t)) =& (1 - \alpha_t) q_{i, t} (v(t), c_i(t)) \nonumber \\
	& + \alpha_t [ r_i(t) + \gamma \max_{c'_i} q_{i,t}(v(t+1), c'_i)], \label{Eq:update-state}
\end{align}
where $v(t+1)$ is the next state. The learning rate parameter $\alpha_t$ is crucial and ranges from $0$ to $1$. A smaller $\alpha_t$ makes the learning progression persistent and effective, while the algorithm with a larger $\alpha_t$ would forget the information learned in the past too rapidly. 

In the single-agent setting, there are theoretical guarantees on the convergence; see \cite{watkins1992q} for classic Q-learning and \cite{fan2020theoretical} for deep Q-learning. However, in multi-agent Q-learning, when other agents are part of state variables, the environment becomes non-stationary in the eyes of each agent. Competitors' actions change the state since they are components of the LOB. An agent's policy depends on the state and therefore his rivals' policies, which are also changing over periods by learning or experimenting under $\varepsilon$-greedy policies. Therefore, multi-agent Q-learning currently lacks general convergence guarantees, due to the technical difficulties of non-stationarity. In practice, convergence is verified only ex-post. In this paper, we adopt a stopping criterion as the greedy policy does not change for $10^5$ consecutive periods. The algorithm converges practically, thanks to the relatively simple economic environment with low-dimensional state and action spaces. 

\section{Experiments}\label{sec:exp}
\subsection{Baseline model}\label{sec:baseline}
We first consider an exchange environment with two market makers. Set the number of price levels on one side as $K=4$. $\delta_k = k$ is interpreted as the multiples of one tick size. The set of feasible actions contains the specification of probability mass on each price level. Suppose ask or bid action concentrates mainly on one level and quotes others with relatively small percentages. In the baseline model, ask or bid action $k$ quotes 70\% on level $k$ and 10\% on each level $j \neq k$, respectively. The joint (ask, bid) actions are encoded in row-major order. For example, joint action 5 is (ask level 2, bid level 1). For simplicity, assume each maker quotes a fixed volume of $20$ limit orders on one side.      

To summarize the market condition as a state variable, suppose the market makers calculate the weighted average of spreads on each side and encode it as a discrete state. Since it is based on the LOB, the state variable relies on agents' aggregated quotes distributions, but not on each agent's actions directly. 

There are many choices for the weights $w_k$ in the arrival probability \eqref{eq:arrprob}. To make the arrival probability decrease rapidly, we propose to consider a polynomial function growing on $\delta_k$:
\begin{equation}\label{eq:weight}
	w_k = c_1 (\max\{\delta_k + \eta - c_0, 0\})^2.
\end{equation}
The maximum operator in \eqref{eq:weight} handles the negative $\eta$ in taker-maker models discussed later. With $c_1=0.2, \eta=0, c_0 = 1$ in \eqref{eq:weight}, the weights are given by $\{0.00,  0.20,  0.80,  1.80\}$. Let the mid-price volatility $\sigma = 0.4$. These values are arbitrarily chosen. Alternatively, one can fit the model with market data. High price levels are unlikely to receive many market orders. Besides, we specify the largest market order size $M$ as a constant equal to the volume available in the LOB.  

Set the inventory risk aversion constant $\xi=0.05$. Figure \ref{fig:reward} plots the one-period reward for one agent under all action pairs. The joint action 6, or (ask 2, bid 2), achieves the highest reward of $34.2$. It is also not optimal to deviate from action 6. By Definition \ref{def:nash} and \ref{def:coop}, action 6 is both the Nash equilibrium and the cooperative strategy for the one-period game. The lowest ask and bid level, with a reward of $26.5$, is not a Nash equilibrium, since the reward is higher if we choose action 6 instead when the competitor selects action 1.
\begin{figure}
	\centering
	\includegraphics[width=0.9\linewidth]{reward.pdf}
	\caption{Reward matrix for one agent}
	\label{fig:reward}
\end{figure}

We set constant learning rate $\alpha=0.05$ and exploration rate $\mu = 10^{-5}$ in the baseline setting. For the initialization of Q-values, \cite{sutton1996generalization,zhang2017deeper} documented that zero initial Q-values encourage exploration, which is adopted as the default method.


\begin{figure}
	\centering
	\includegraphics[width=0.7\linewidth]{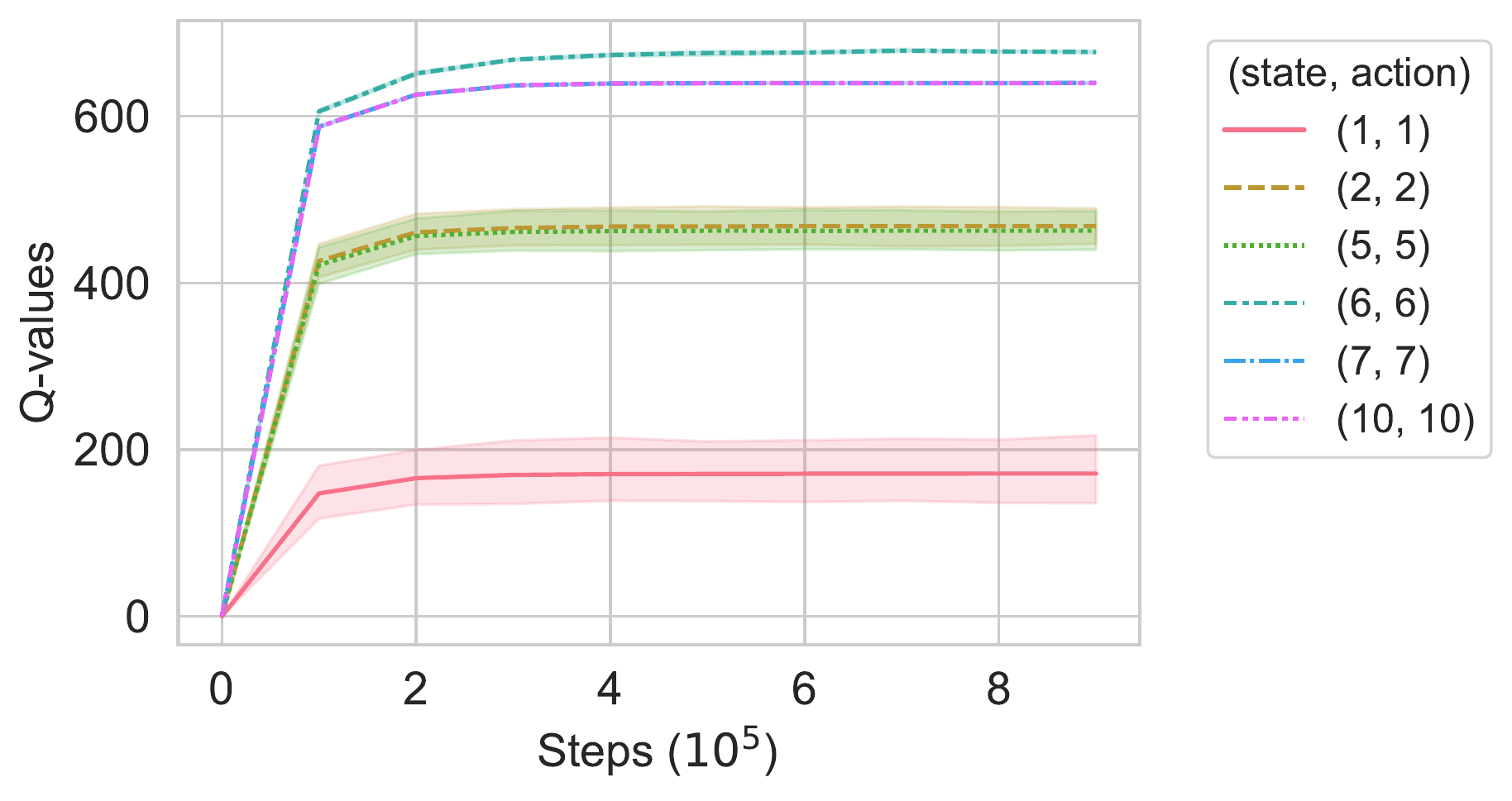}
	\caption{$Q$-values}
	\label{fig:QConverge}
\end{figure}

\begin{figure}
	\centering
	\includegraphics[width=0.7\linewidth]{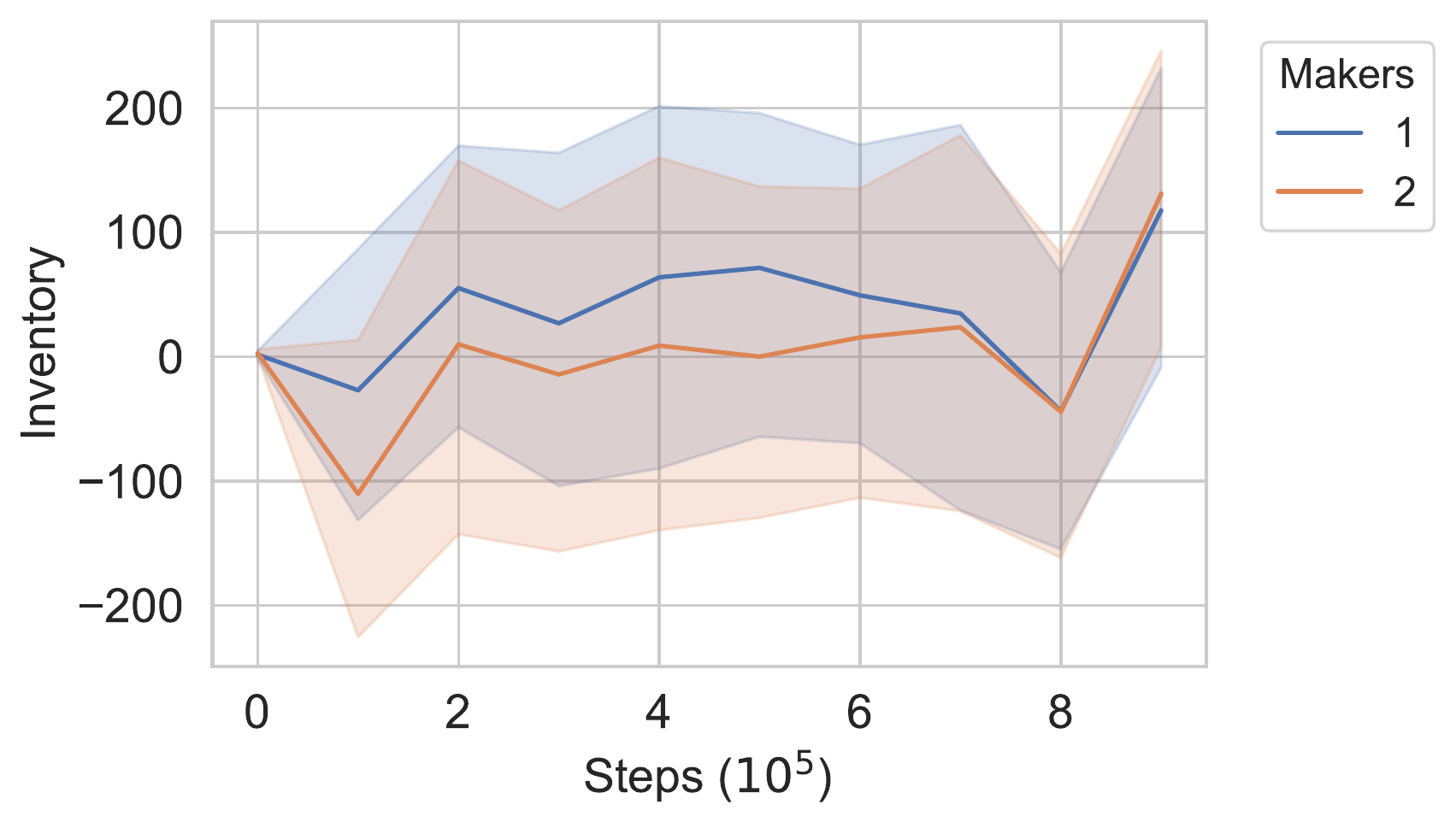}
	\caption{Inventory}
	\label{fig:Inventory}
\end{figure}

For each parameter setting, we repeat the program for 20 instances. Take the no-maker-taker fee setting with $\gamma=0.95$ as an example. Figure \ref{fig:QConverge} illustrates that the algorithm converges in the long run. The solid curve and the shadowed area are the mean values and confidence intervals, respectively. For simplicity, we only report several large $Q$-values in certain action-state pairs. Although the theoretical convergence guarantee is open for multi-agent reinforcement learning with states, the experimental results in Figure \ref{fig:QConverge} give strong evidence of convergence. Figure \ref{fig:Inventory} demonstrates the effectiveness of inventory control. Besides the inventory risk aversion in the payoff, the agent also implements the skewing. When the inventory is higher than an upper bound, the maker publishes more attractive ask prices than the bid ones. In the experiment of Figure \ref{fig:Inventory}, we suppose agents consider $-500$ and $500$ as thresholds. The inventories of agents remain low during the entire learning progression. 

\subsection{Maker-taker fees}
Table \ref{tab:maketake} reports the mean values and standard deviations in parentheses of net fees and transactions per agent and per period after convergence. The net fees are for one side and equal to $\delta + \beta$. In the experiment, we suppose a market maker quotes limit orders with a total volume of 20 on one side. The distribution is specified in the way explained in Section \ref{sec:baseline}. Market order volume on one side is at most 20 multiplied by the number of agents. The last column of Table \ref{tab:maketake} shows orders received per agent on one side.     

\begin{figure}
	\centering
	\includegraphics[width=0.7\linewidth]{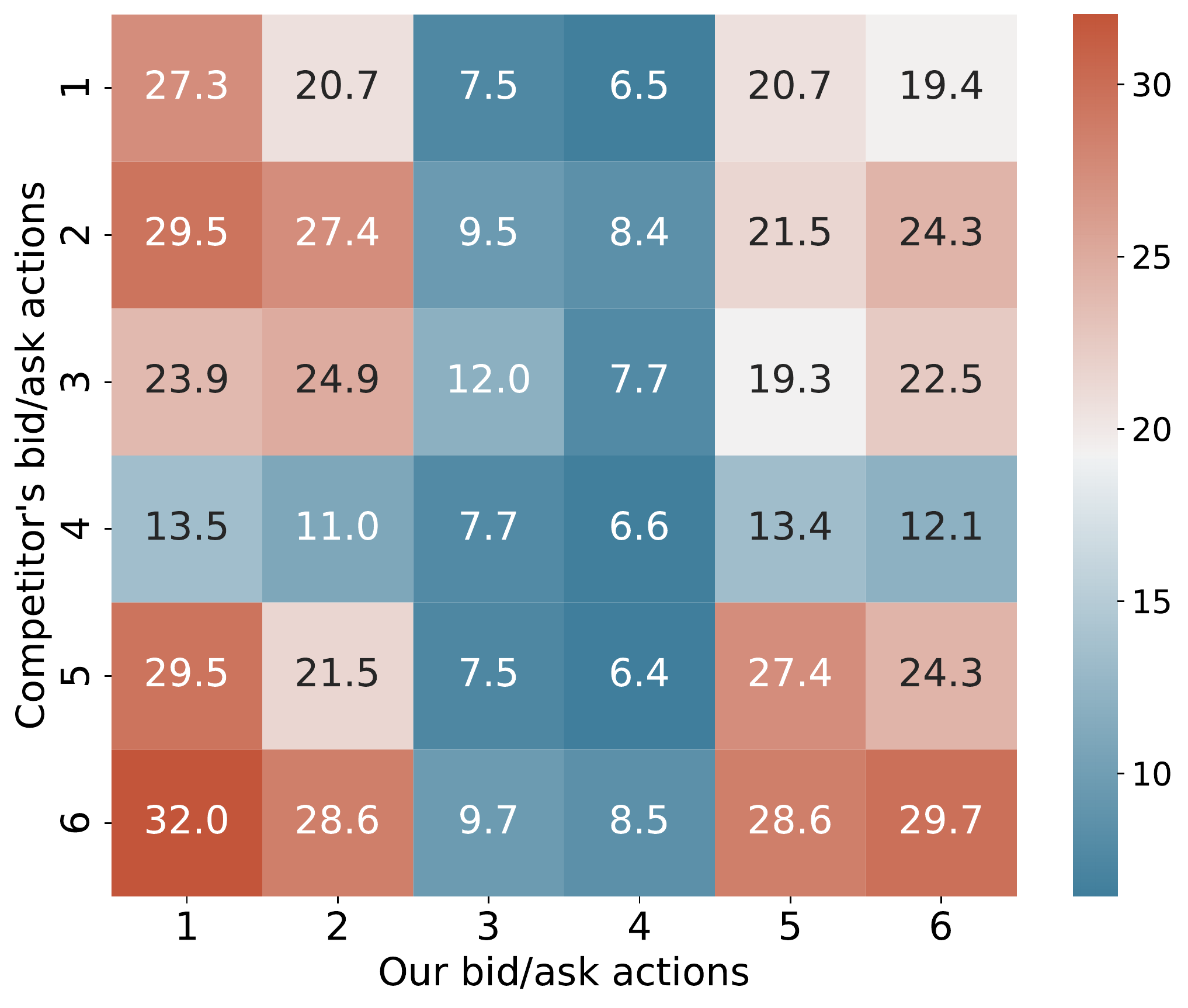}
	\caption{Part of reward matrix with $\beta=0.1$}
	\label{fig:beta01reward}
\end{figure}

\begin{figure}
	\centering
	\includegraphics[width=0.7\linewidth]{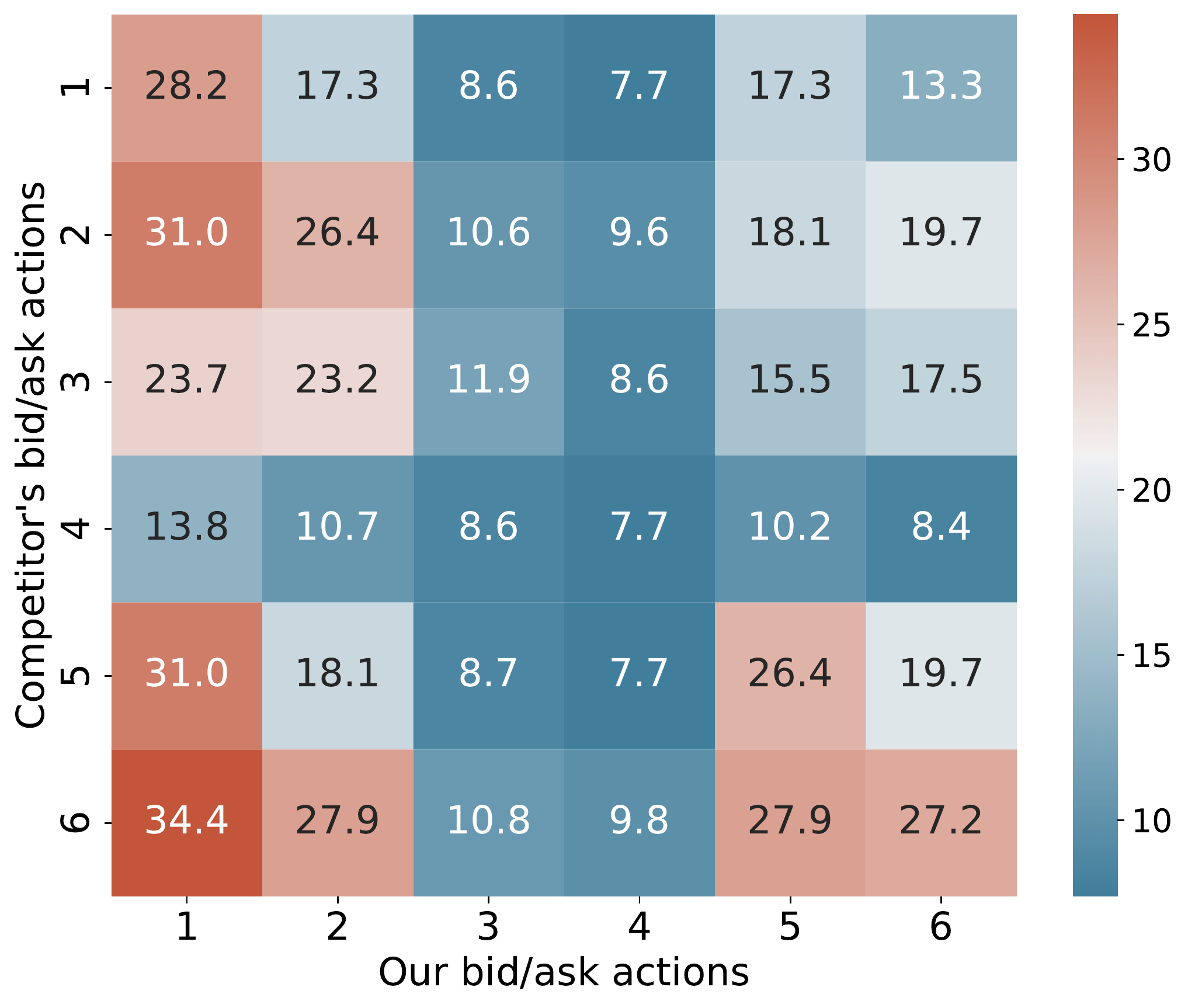}
	\caption{Part of reward matrix with $\beta=0.2$}
	\label{fig:beta02reward}
\end{figure}

Consider a discount factor $\gamma = 0.95$ first. When there is no maker-taker fee, Table \ref{tab:maketake} indicates that the algorithm converges to the joint action 6 with (ask 2, bid 2), which is both the Nash equilibrium and cooperative strategy. It is in line with the theoretical reward matrix in Figure \ref{fig:reward}. Next, the exchange adopts the maker-taker fee model. We keep the difference between the access fee and the rebate at 0.05, i.e., $\eta = \beta + 0.05$. With $\beta=0.1$, or equivalently 10\% of the tick size, Figure \ref{fig:beta01reward} gives part of the reward matrix while omitting higher spreads for simplicity. The rewards include rebates. The joint action 6 is no longer the Nash equilibrium. When our competitor chooses action 6, it becomes more profitable to deviate and select action 1. However, action 6 is still the cooperative strategy with the highest reward. In contrast to the theoretical results, the algorithm does not converge to the Nash equilibrium and still charges action 6 in more instances than action 1. In our experiments, 80\% of all instances converge to the joint action 6. When the rebate further increases as in Figure \ref{fig:beta02reward} to $\beta=0.2$, it becomes more profitable to undercut the competitor, as the reward for the action pair (ours=1, competitor's=6) increases to 34.4 from 32.0 in Figure \ref{fig:beta01reward}. Experimentally, we achieve a lower transaction cost in Table \ref{tab:maketake} when $\gamma = 0.95$. However, the cost is not always decreasing for intuitive reasons. When the agents have quoted the lowest level 1, a further increment merely increases the transaction costs of customers and benefits the market makers.
\begin{table}
	\centering
	\begin{tabular}{c c c c}\toprule
		discount factor      &  rebate             & net fee             & order  \\ \midrule
		$\gamma = 0.00$  & $\beta = 0.0$  & 1.648 (0.356) & 10.178 (2.043) \\
		& $\beta = 0.1$  & 1.112 (0.055) & 12.412 (1.103) \\ \midrule
		$\gamma = 0.50$  & $\beta = 0.0$  & 1.765 (0.306)  & 9.654 (1.838) \\
		& $\beta = 0.1$  & 1.112 (0.056)  & 12.410 (1.109)  \\ \midrule
		$\gamma = 0.95$  & $\beta = 0.0$  & 1.887 (0.040)  & 9.032 (1.142) \\
		& $\beta = 0.1$  & 1.792 (0.342)  & 8.517 (2.228)  \\
		& $\beta = 0.2$   & 1.207 (0.053)  & 11.741 (1.130) \\
		& $\beta = 0.3$   & 1.304 (0.046)  & 10.996 (1.141) \\ \midrule
		$\gamma = 0.99$ & $\beta = 0.0$  & 1.887 (0.037) &  9.031(1.140) \\
		& $\beta = 0.1$  &  1.948 (0.092) &  7.543 (1.218) \\
		& $\beta = 0.2$   &  2.040 (0.044) & 6.612 (1.078) \\
		& $\beta = 0.5$   & 1.523 (0.095) & 9.132 (1.443) \\ \bottomrule
	\end{tabular}
	\caption{Maker-taker model}
	\label{tab:maketake}
\end{table}

For the multi-period game, a large discount factor $\gamma$ enables the agent to be far-sighted, which leads to higher profits in the long run. When we vary the discount factor $\gamma$, Table \ref{tab:maketake} shows that the rebate $\beta = 0.1$ reduces the transaction costs in the short-sighted setting but increases the costs in the far-sighted case. Market makers can earn even more profits than the no-rebate case under $\gamma = 0.99$ and $\beta=0.1$. Although an upper bound on maker-taker fees is more common in practice, our experiments advocate that we may need a lower bound for the rebate to make it attractive and thus effective. Besides other arguments on the suitable level of maker-taker fees, algorithmic cooperation provides another perspective to investigate this long-standing problem.   

For simplicity, we only consider $\gamma = 0.95$ in the following subsections.

\subsection{Taker-maker fees}
There are also some exchanges with a ``taker-maker'' fee model charging makers a fee and paying takers a rebate. According to the appeal mentioned in the introduction, four out of thirteen national exchanges have a taker-maker model. In contrast, seven exchanges adopt the maker-taker model. Two others charge a flat (or no) fee and offer no rebates. 

We implement a taker-maker model by a negative ``rebate'' paid to the maker. The weight function is the same as in \eqref{eq:weight} and $\eta = \beta + 0.05$. Note that when $\beta < 0$ and $\eta<0$, the weight for $\delta=1$ is $w_1 = 0$ by $c_1=0.2$ and $c_0 = 1$. In other words, a further improvement in compensation will not make the lowest spread more attractive since the arrival probability is one already. Thus, the compensation mainly attracts more orders for higher levels. Table \ref{tab:takemake} contains experimental results with three rebate levels. Compared with the maker-taker model in Table \ref{tab:maketake}, the transaction fees reduction is from compensations to investors instead of cutting off spreads by market makers. Note that each pair of compensation and net fee in Table \ref{tab:takemake} implies the spread level is roughly equal to 2. Besides, the taker-maker model is less effective than the maker-taker model under the current formulation. It needs a much higher level $\eta=-0.75+0.05=-0.7$ to achieve a similar net fee in the maker-taker model with $\beta=0.2$. Nevertheless, the advantage of a taker-maker model is that the trading volume increases significantly, thanks to the stimulus of compensation to investors.

\begin{table}
	\centering
	\begin{tabular}{c c c}\toprule
		rebate & net fee & order \\ \midrule
		$\beta = -0.15$ & 1.751 (0.044)   & 10.045 (1.137)  \\
		$\beta = -0.45$ & 1.479 (0.039) & 13.101 (1.074) \\
		$\beta = -0.75$ & 1.254 (0.056)  & 15.690 (0.927) \\ \bottomrule
	\end{tabular}
	\caption{Taker-maker model}
	\label{tab:takemake}
\end{table}

\subsection{The impact of volatility}
From the design of order arrival probability in \eqref{eq:arrprob}, high mid-price volatility implies a strong trading willingness of investors. In the case with $\sigma \rightarrow \infty$, the probability converges to one for all price levels. It facilitates the cooperation between makers to maintain wide spreads. Table \ref{tab:vol} confirms this analysis as it shows that net fees and market order volumes climb up with volatility.

Consider the rebate level $\beta = 0.2$ in this subsection, which reduces transaction fees in the baseline setting. Table \ref{tab:vol} gives that low volatility of 0.2 can stabilize the lowest spread as the long-run outcome. The variance in net fees is extremely low. However, trading volumes are small, mainly due to the low trading willingness of investors. It becomes unneeded to provide rebates $\beta = 0.2$ when the mid-price is stable. We can reduce the rebate while maintaining the lowest spread level.   

When mid-price volatility increases, the rebate becomes less effective since makers realize that higher spreads can also attract a lot of market orders. Table \ref{tab:vol} indicates that makers switch to the level $2$ under a high volatility of $0.6$. Therefore, we should consider charging more taker fees to make high spreads less attractive. However, we may face a dilemma that the required taker fee for cutting down spreads is too high. Then the net transaction costs are even close to the levels without rebates. For such cases, the maker-taker model becomes useless in preventing cooperation between makers.                  
\begin{table}[!t]
	\centering
	\begin{tabular}{c c c}\toprule
		volatility & net fee &  order  \\ \midrule
		$\sigma=0.2$ & 1.200 ($2.22 \times 10^{-16}$)   & 6.905 (1.093)  \\
		$\sigma=0.6$ & 2.095 (0.043) & 9.573 (1.133) \\
		$\sigma=1.0$ & 2.127 (0.044) & 12.848 (1.081) \\ \bottomrule
	\end{tabular}
	\caption{The impact of volatility}
	\label{tab:vol}
\end{table}

\subsection{The impact of inventory risk}
\begin{figure}
	\centering
	\includegraphics[width=0.7\linewidth]{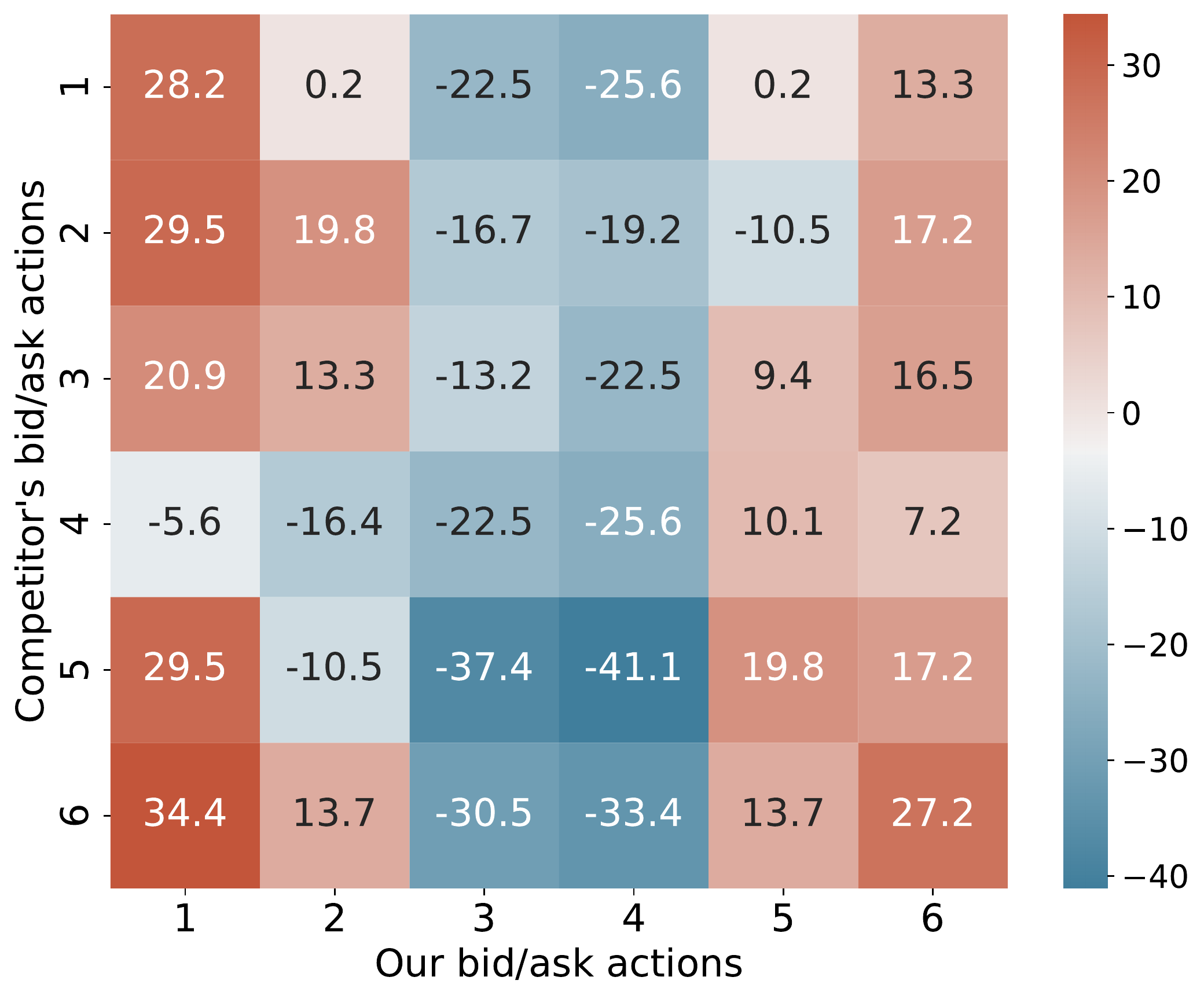}
	\caption{Part of reward matrix with inventory aversion coefficient $\xi=0.3$}
	\label{fig:inv03}
\end{figure}

Inventory risk appears in two aspects of our market-making model. First, it alters the reward mechanism through the aversion coefficient $\xi$. Actions that receive potential imbalanced bid-ask market orders are less favored. Figure \ref{fig:inv03} presents a corner of the reward matrix with actions $1 \sim 6$, inventory risk aversion coefficient $\xi=0.3$, and the market rebate $\beta = 0.2$. Action 3 (ask=1, bid=3) and 4 (ask=1, bid=4) are highly imbalanced and lead to negative payoffs. Second, makers may skew their quotes to offset the inventory actively. In this subsection, we still adopt the skewing method introduced in the baseline model. 

We vary the aversion coefficient $\xi$ and summarize the results in Table \ref{tab:inv}, under the rebate $\beta = 0.2$ and taker fee $\eta = 0.25$. $\xi=0.0$ corresponds to the situation that agents are insensitive to the impact of inventory on rewards. When makers are more averse to inventory risk, the last column of Table \ref{tab:inv} indicates that they are less willing to attract market orders. The trade per period and per agent decreases to $7.132$ from $11.484$. Implicitly, makers may prefer higher spreads since these choices are less likely to receive market orders and thus, induce fewer inventory changes. Another potential reason is in Figure \ref{fig:inv03}. When both agents quote action 6 and one wants to undercut the spreads, the best choice is action 1 with the lowest level on both sides. There are also imbalanced actions $2 \sim 5$ which use the lowest spread only on one side. But under a high inventory risk aversion, the payoffs for these actions are too low and unfavorable, as shown in Figure \ref{fig:inv03}. It leads to a smaller chance of hitting the lowest spread on one side. Eventually, action 6 is more stable and becomes the long-run outcome for most sessions with $\xi = 0.3$. 

In summary, when makers are more averse to inventory risk, the rebates should be higher to reduce transaction costs. These rebates serve as compensation to encourage makers to bear the inventory risk and provide liquidity. Nevertheless, it may be hard to evaluate the inventory aversion levels of agents and adjust the rebates accordingly. In practice, regulators may focus on the outcomes directly to select the fee levels.        

\begin{table}[H]
	\centering
	\begin{tabular}{c c c}\toprule
		aversion &  net fee & order  \\ \midrule
		$\xi = 0.0$  & 1.248 (0.134) & 11.484 (1.359) \\
		$\xi = 0.1$ & 1.206 (0.050) & 11.740 (1.134) \\
		$\xi = 0.2$ & 1.811 (0.374) & 8.031 (2.528) \\
		$\xi = 0.3$ & 1.958 (0.252) & 7.132 (1.868) \\ \bottomrule
	\end{tabular}
	\caption{The effect of inventory risk aversion}
	\label{tab:inv}
\end{table}

\subsection{Number of agents}
A pearl of common wisdom is that cooperation between more agents is more difficult. In Table \ref{tab:agent}, we increase the number of agents while keeping other parameters unchanged. Set the maker-taker fees as zero to focus on the impact of the number of agents. When there are more makers, we suppose market orders also increase proportionally and follow $\text{binomial}(20N, p)$, where $N$ is the number of agents and $p$ is given in \eqref{eq:arrprob}. The motivation is that higher LOB volumes can attract more investors. Statistics for the four-agent case are very close to the two-agent one in Table \ref{tab:maketake}. The order received per agent is approximately $9.08$. However, when there are enough market makers, Table \ref{tab:agent} indicates that the long-run spread converges to the lowest level. Moreover, the last column in Table \ref{tab:agent} shows that orders received per agent increase to $13.25$. Market quality is improved significantly with more makers.

For a trading venue with enough makers, maker-taker fees will increase transaction costs for investors instead of promoting competition between makers. The reason is that competition is already intense enough to reach the lowest quote level. Higher rebates have no benefit on fee reductions. It suggests that regulators have to include the number of agents as a factor when evaluating the effectiveness of maker-taker fees.   
\begin{table}
	\centering
	\begin{tabular}{c c c}\toprule
		Number of agents & net fee & order  \\ \midrule
		4 & 1.877 (0.051)  & 9.075 (0.808)  \\
		6 & 1.061 (0.152)  & 12.965 (0.978) \\ 
		8 & 1.004 (0.012)  & 13.249 (0.529) \\ \bottomrule
	\end{tabular}
	\caption{The effect of number of agents}
	\label{tab:agent}
\end{table}

\section{Conclusion}\label{sec:con}
This paper utilizes an experimental approach to understand the role of maker-taker fees in algorithmic cooperation. Indeed, our results hold under some idealized assumptions that could limit generalization. The debate on maker-taker fees is likely to continue. A future direction is to consider more realistic settings and create innovative approaches to calibrate the model with empirical data. For example, we may extend to the continuous-time setting. The number of market makers may also vary with time. Another crucial and hard open question is theoretical guarantees on the convergence to cooperative or competitive strategies.

%
\bibliographystyle{ACM-Reference-Format}
\bibliography{reference}


\end{document}